\documentclass[aps,pre,twocolumn,superscriptaddress]{revtex4-1}
\usepackage{mathtools}
\usepackage{amssymb,amsmath}
\usepackage{graphicx}
\usepackage{algorithmicx}
\usepackage{algpseudocode}
\usepackage{algcompatible}
\usepackage{algorithm}
\usepackage{soul, color}
\usepackage{csquotes}
\usepackage{enumitem}
\makeatletter
\newenvironment{breakablealgorithm}
  {
   \begin{center}
     \refstepcounter{algorithm}
     \hrule height.8pt depth0pt \kern2pt
     \renewcommand{\caption}[2][\relax]{
       {\raggedright\textbf{\ALG@name~\thealgorithm} ##2\par}%
       \ifx\relax##1\relax 
         \addcontentsline{loa}{algorithm}{\protect\numberline{\thealgorithm}##2}%
       \else 
         \addcontentsline{loa}{algorithm}{\protect\numberline{\thealgorithm}##1}%
       \fi
       \kern2pt\hrule\kern2pt
     }
  }{
     \kern2pt\hrule\relax
   \end{center}
  }
\makeatother

\sethlcolor{cyan}

\newcommand{\fref}[1]{Fig.~\ref{fig:#1}}

\newcommand{\flabel}[1]{\label{fig:#1}}
\newcommand{\eref}[1]{Eq.~\ref{eqn:#1}}

\newcommand{\elabel}[1]{\label{eqn:#1}}

\begin{document}

\title{Multiscale simulations of anisotropic particles combining  Brownian Dynamics and Green's Function Reaction
    Dynamics} 
  
  \author{Adithya Vijaykumar}
  \affiliation{FOM Institute AMOLF, Science Park 104, 1098 XG Amsterdam, The Netherlands}
  \affiliation{van 't Hoff Institute for Molecular Sciences, University of Amsterdam, PO Box 94157, 1090 GD Amsterdam, The Netherlands}

  \author{Thomas E. Ouldridge}
  \affiliation{Department of Bioengineering,
Imperial College London, South Kensington Campus, London SW7 2AZ, United Kingdom}
  
  \author{Pieter Rein ten Wolde}
  \email[]{Electronic mail: p.t.wolde@amolf.nl}
  \affiliation{FOM Institute AMOLF, Science Park 104, 1098 XG Amsterdam, The Netherlands}

  \author{Peter G. Bolhuis}
 \email[]{Electronic mail: p.g.bolhuis@uva.nl}
  \affiliation{van 't Hoff Institute for Molecular Sciences, University of Amsterdam, PO Box 94157, 1090 GD Amsterdam, The Netherlands}

\begin{abstract}

The modeling of  complex reaction-diffusion processes in, for instance, cellular biochemical networks or self-assembling  soft matter can be tremendously sped up by employing a multiscale algorithm which 
combines the mesoscopic Green’s Function Reaction Dynamics (GFRD) method with explicit stochastic Brownian, Langevin, or deterministic Molecular Dynamics to treat reactants at the microscopic scale
[A. Vijaykumar, P.G. Bolhuis and P.R. ten Wolde, J. Chem. Phys. {\bf 43}, 21: 214102 (2015)]. Here we extend this multiscale BD-GFRD approach to include the orientational dynamics that is crucial to describe the anisotropic interactions often prevalent in biomolecular systems. We illustrate the novel algorithm using a simple patchy particle model. After validation of the algorithm we discuss its performance. The rotational BD-GFRD multiscale method will open up the possibility for large scale  simulations of e.g. protein signalling networks.
\end{abstract}
  \maketitle



\section{Introduction}

Complex systems such as biochemical networks in living cells,
catalytic reactions in, e.g.~a fuel cell, 
surfactant/water/oil mixtures, or self-assembling soft 
matter, can be modeled efficiently as reaction-diffusion systems.  In
such reaction-diffusion systems the spatial distribution of reactants
and the stochastic nature of their interactions are crucial for the
system's macroscopic behaviour. At sufficiently low
concentrations, the time taken for the reactants to diffuse and
randomly find each other is much larger than the time required for the
reaction. For example, in cellular systems, the concentrations of
  proteins are often in the ${\rm nM-\mu M}$ range and their diffusion
  constants in the $1-10\mu{\rm m}^2{\rm s}^{-1}$ range. This means
  that, with typical protein cross sections of $10 {\rm nm}$, the time
  it takes for reactants to find each other is on the order of milliseconds to
  seconds. This is often much longer than the microsecond timescales
  on which the actual  association events occur
  once the particles have found each other
\cite{Schreiber:2009cd,tenWolde:2016ih}. Reaction-diffusion
  systems thus often exhibit a strong separation of length and time
  scales, with the diffusive search process happening on length and
  timescales of microns and milliseconds to seconds, and the reactions
  occurring on scales of nanometers and sub-milliseconds
\cite{tenWolde:2016ih}. Simulating such systems with conventional,
  brute-force simulation techniques is notoriously difficult. Indeed,
  simulating cellular biochemical networks with straightforward
  brute-force Brownian Dynamics (BD)
\cite{Andrews2004,Lipkova2011,Flegg2012,Schoneberg:2013ek} often
  means that most CPU time is spent on propagating the particles
  towards one another \cite{vanZon:2005vt}.  To overcome the
inefficiency of straightforward BD requires special techniques such as
Green's Function Reaction Dynamics
(GFRD)\cite{vanZon:2005cy,Takahashi2010}.

GFRD is a mesoscopic technique that decomposes the many particle
reaction diffusion problem into sets of one- and two-body problems that
can be solved analytically. This is achieved by putting single
particles and pairs of particles in so-called  \emph{protective} domains
that do not overlap with each other. For each of these domains the
reaction-diffusion problem is solved analytically using Green's
  functions. This yields for each domain a next event $\it{type}$
which can either be a reaction in the domain or an escape from the
domain, as well as a next event $\it{time}$, i.e.~the time at which
this event occurs. These events are put in a scheduler list which is
updated chronologically. This makes GFRD an asynchronous, event-driven
algorithm. Since stochastic processes in the individual domains are
independent of each other, GFRD is an exact algorithm to simulate
large reaction-diffusion systems.  As the particles make huge leaps
in space and time in GFRD the computational effort in propagating the
particles to one another is greatly reduced, making GFRD orders of
magnitude faster than brute force BD. 
However, the particles are assumed to be idealized spheres interacting
via an isotropic potential  and the reactions to occur according
to intrinsic rates in pair domains.
Solving the Green's function for reactive events
involving the complex anisotropic potentials required for proper
modeling of proteins or other molecules is extremely cumbersome, and
in fact most likely will reduce the efficiency of the GFRD approach
substantially.  In contrast, straightforward BD is able to naturally
simulate orientational dynamics of protein particles with complex
anisotropic (effective) interactions.

This observation raises the question whether it is possible to combine
the computational power of GFRD with the microscopic detail of BD.  In
previous work, we introduced a novel multi-scale scheme, called
  Molecular Dynamics-GFRD (MD-GFRD), which
combines GFRD with a microscopic simulation technique such as
deterministic molecular dynamics (MD), or stochastic Langevin Dynamics or 
Brownian Dynamics (BD)\cite{Vijaykumar2015}.  In this scheme GFRD
handles diffusion of particles at the mesoscopic scale, while MD, LD or BD
treats the particles that are coming close to each other. In 
previous work and here, we limit ourselves to BD, although the scheme
can very easily formulated for MD and LD. The
multi-scale algorithm defines the micro- and mesoscopic regions
adaptively on the fly and switches seamlessly between the two
techniques based on predefined scenarios. 

In this work we extend MD-GFRD to incorporate the orientational
  dynamics of particles that interact via an anisotropic potential. As
  in the original MD-GFRD technique \cite{Vijaykumar2015}, GFRD
  is used for propagating the particles towards one another when they
  are far apart. Once the particles are within a predefined threshold
  distance from each other, the algorithm switches to BD. The
  complex orientational dynamics once the particles are close together
  is thus simulated with BD. When the particles are bound, MD-GFRD
  could in principle continue to simulate these particles with BD. However, in many cases, and typically in cellular systems, the
  particles are bound much longer than the time it takes to diffuse
  and thermalise
  within the interaction well, meaning that dissociation is a rare
  event.  MD-GFRD exploits this separation of timescales by treating
  the dissociation as a first order reaction, with an intrinsic dissociation rate constant
  that has been pre-determined. After dissociation, the
  particles can be propagated again with GFRD. Importantly, however,
  after dissociation the particles do not immediately loose their
  orientational memory, which means that they must be propagated with
  Green's Functions that do not only describe the translational
  dynamics of the particles, but also their orientational dynamics. In
  this paper, we describe in detail how the MD-GFRD scheme switches
  between MD and GFRD and how this switching depends on the
  translational and orientational dynamics of the particles. We also
  present the Green's Functions that allow GFRD to simulate the
  particles' orientational dynamics.

  The remainder of the paper is organized as follows.  In the methods
  section we first give an overview of the MD-GFRD algorithm. Then we
  describe how the algorithm simulates the diffusion of particles with
  rotational degrees of freedom, both for particles in BD and GFRD
  mode.  We discuss how MD-GFRD handles the association-dissociation
  reactions, and we describe how it switches between BD and GFRD
  propagation.  In many systems, including that studied here,
  dissociation is a rare event. This means that computing the intrinsic
  dissociation rate constant, as used by MD-GFRD, requires rare event
  methodology, like Transition Interface Sampling \cite{vanErp2003}
  and Forward Flux Sampling (FFS) \cite{Allen:2005dn}. Here, we
  briefly describe how we use FFS to pre-compute the dissociation rate
  constant.  We then illustrate the new technique by simulating the
  association and dissociation of patchy particles.  In many cases,
  globular proteins can be coarse-grained as so-called \emph{patchy
  particles}, where the complex binding sites are modeled as patches
  on a spherical particle.  These patchy particles also play an
  important role in the modeling of soft matter\cite{Glotzer:2007,Newton:2016}.
We demonstrate that the algorithm
reproduces quantities that can be obtained analytically such as the
equilibrium constants, binding probabilities and the power spectra of
the binding reactions. We end with a discussion of the performance
of the algorithm.


\section{Methods}
\subsection{Summary of multiscale approach}
\label{sec:summary}
The MD-GFRD algorithm is a generic algorithm that enables
simulation of any reaction-diffusion system at the particle
  level. It allows for mono-molecular reactions of the type ${\rm A}
  \to {\rm B} + {\rm C} + \dots$ and bi-molecular reactions of the type
  ${\rm A}+{\rm B} \to {\rm C}+{\rm D}+\dots$. By combining these two
  reactions, any complex biochemical network can be simulated. Here,
  however, we will limit ourselves to simple association-dissociation
  reactions ${\rm A}+{\rm B} \leftrightarrows {\rm C}$.
\begin{figure}[b]
\includegraphics[width=8.3cm]{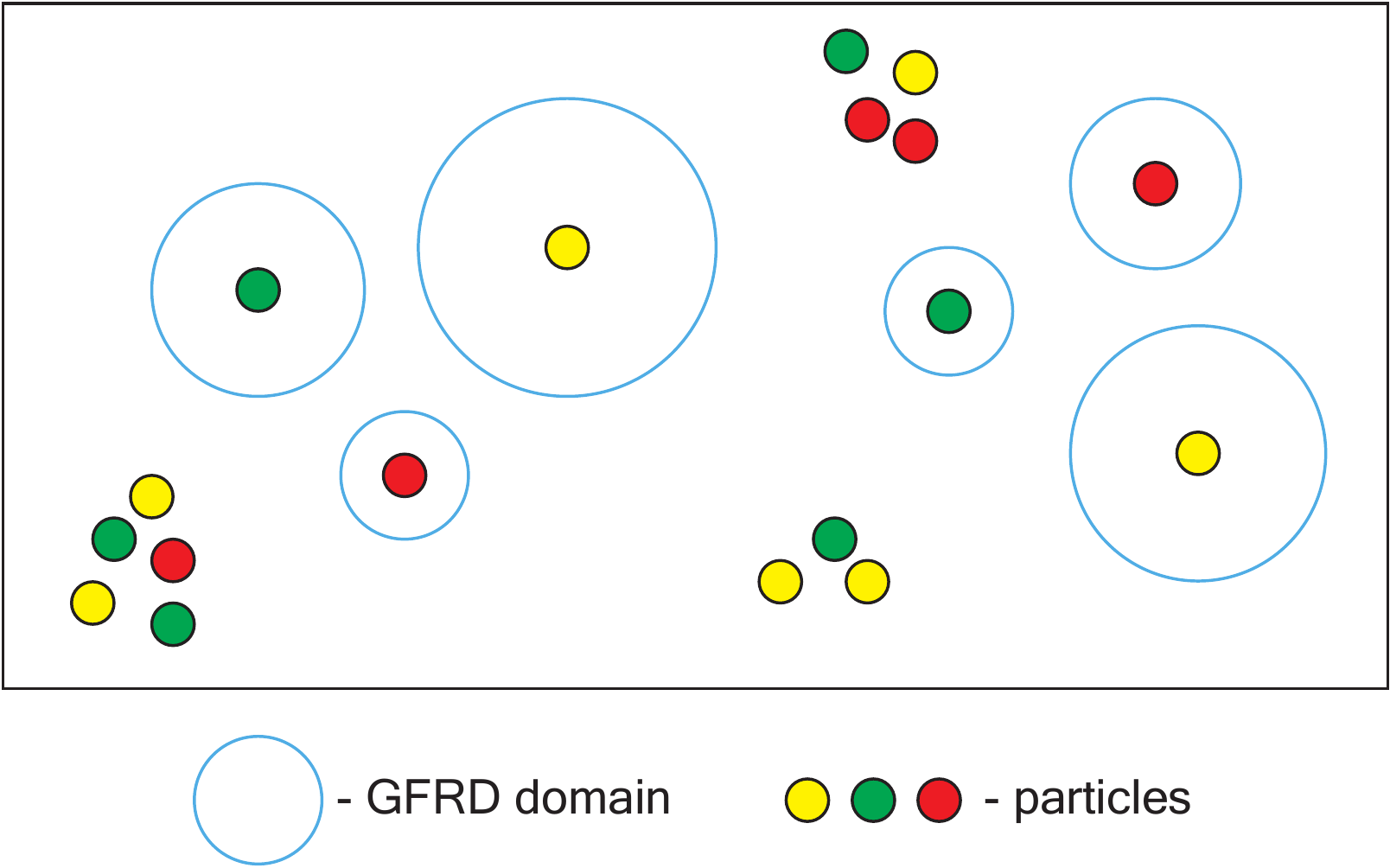}
\caption{\flabel{fig:overview} 
MD-GFRD scheme: particles that are far away other particles are put into a GFRD 
domain. For each GFRD domain the next event \textit{time} and \textit{type} is determined. 
These next event times are added to a chronologically ordered event list, 
and updated when the simulation time has reached the time of the next event. 
Particles that are close to other particles are propagated collectively with Brownian Dynamics.}
\end{figure}
  The MD-GFRD algorithm distinguishes two types of particles as shown in \fref{fig:overview}: 1) BD
  particles that are propagated collectively in a conventional,
  brute-force manner using small time steps, and 2) GFRD particles
  that are updated asynchronously in an event-driven manner 
Single particles that are
  sufficiently far away from all other particles according to a predefined
  cut-off distance are put into protective domains. For each of these
  domains, the algorithm determines, as in the conventional GFRD
  scheme \cite{Takahashi2010}, the next-event {\em type}, which is
  either a mono-molecular decay reaction (such as dissociation) or an
  exit of the particle from the domain, and the corresponding
  next-event {\em time}, which is when this next event will
  happen. The next-event times of the respective GFRD domains are put
  in a chronologically ordered event list, which is updated only when
  the simulation time has reached the time of the first next event.
  The event-driven nature of GFRD allows MD-GFRD to make large jumps in
  space and time when the domains are large. It is the origin of the
  high efficiency of the scheme.

  The other particles are simulated explicitly with BD. This part of
  the algorithm takes into account the forces between the particles
  when they come within the interaction range of the potential from
  each other. The BD propagation also naturally simulates the
  association reaction ${\rm A}+{\rm B} \to {\rm C}$: two particles A
  and B form the bound complex C when they enter the well of the
  interaction potential. The two monomers A and B in the dimer C could
  be propagated separately with BD, but it is more efficient to
  propagate them as a single particle C. The dissociation of C into A
  and B is then treated as a uni-molecular reaction event, which is
  added to the event list.

  BD propagation is continued until one of the following events
  occurs: $i$) the simulation time reaches the time of the first event
  in the event list, the event being the escape of a particle from a
  GFRD domain; $ii$) the simulation time reaches the time of the first event
  in the event list, the event being the decay of a GFRD particle, e.g.~ the
  dissociation of C into A and B; $iii$) a BD particle dissociates into its products, e.g.~ the
  dissociation of C into A and B; $iv$) two BD particles A and B bind each other to form a
  dimer species C; $v$) a BD particle comes too close to a GFRD domain so
  that the GFRD domain must be burst, which means that a position for
  the particle in that domain is generated at the current simulation
  time; $vi$) BD particle moves sufficiently far away from all other BD
  particles and GFRD domains, so that it can be put into a GFRD
  domain. 
 After the event
  has been executed, the system is updated accordingly; for newly
  formed GFRD domains, the next-event types and times are determined
  and inserted into the event list. The propagation of the BD particles is
  then resumed.  The scheme becomes particularly powerful when most
  particles are in GFRD domains. A key objective is thus to keep the
  number of BD particles to a minimum.

The multiscale method that we pursue here involves particles
interacting via anisotropic potentials. This requires an explicit BD
integrator allowing rotational dynamics. Moreover, the GFRD part
requires rotational Green's functions.  In the next subsections we
provide these ingredients, which constitute the most salient
differences of the novel scheme with the previous isotropic MD-GFRD
scheme \cite{Vijaykumar2015}. In the subsequent subsection, we discuss
  in detail how the algorithm switches between GFRD and BD. The next two subsections
  describe how MD-GFRD handles the dissociation events and how the
  dissociation rate constant, needed in MD-GFRD, can be computed efficiently.
In the last subsection, we
  describe the specific interaction potential used to illustrate how
  orientations can be included in  MD-GFRD. \\
  
\subsection{Brownian Dynamics of patchy particles}
Brownian dynamics is used to simulate the solute particles at the microscopic
scales.  In this algorithm the position and the orientation
 of each solute particle in the BD regime is
updated based on the total force and torque acting on the particle.  The force and torque contain a deterministic
component, which arises from the (solvent-mediated) interaction
potential with the other solute particles and the frictional drag from the solvent, and a stochastic
component, originating from the stochastic forces exerted by the solvent
molecules. 
Although the interactions between particles are anisotropic, we model
the particles as spheres of finite radius.
 We represent the rigid body orientation of the particles
using a four component unit vector known as a quaternion, $q=(q_{\rm
  0}, q_{\rm 1}, q_{\rm 2}, q_{\rm 3})$.  The quaternion is an
efficient encoding of the rotation matrix, A given by,
\begin{widetext}
\[ A = \left[ \begin{array}{ccc}
{q_0}^2+{q_1}^2-{q_2}^2-{q_3}^2  & 2(q_1q_2+q_0q_3) & 2(q_1q_3-q_0q_2) \\
2(q_1q_2-q_0q_3) & {q_0}^2-{q_1}^2+{q_2}^2-{q_3}^2 & 2(q_2q_3+q_0q_1) \\
2(q_1q_3+q_0q_2) & 2(q_2q_3-q_0q_1) & {q_0}^2-{q_1}^2-{q_2}^2+{q_3}^2 \end{array} \right].\]
\end{widetext}
which relates vectors in the stationary lab frame, $\hat{u}_{\rm s}$, to the vectors in the moving body 
frame, $\hat{u}_{\rm b}$ via
\begin{equation}
\hat{u}_{\rm s}  = A^T\hat{u}_{\rm b}
\label{eq:spacePatch}
\end{equation}
For example, the vectors $\hat{u}_{\rm b}$  might point to the patches on the surface of the particle which are 
fixed in the body frame.

\begin{figure}[b]
\includegraphics[width=8.3cm]{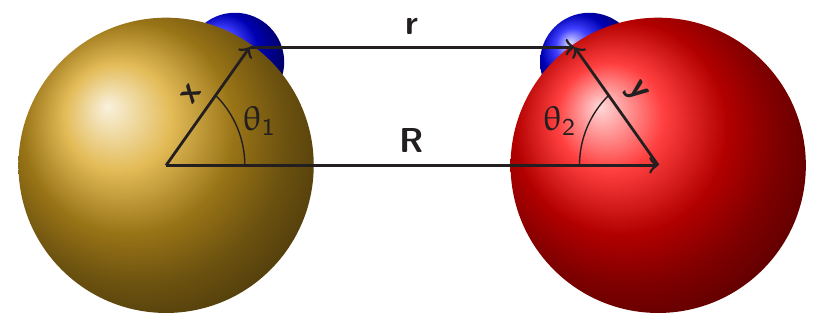}
\caption{\flabel{fig:patchyParticle} 
Each particle may have one or more attractive regions on its surface, called \lq patches',
that facilitate short ranged, highly directional attractive interactions.}
\end{figure}

Each particle has a center of mass, and one or more sticky spots on
its surface called \lq patches' (see \fref{fig:patchyParticle}). 
The particles interact with each other both via a center of mass
isotropic pair potential and via a short ranged isotropic patch-patch
interaction.
We describe the anisotropic model potential that we employ for
illustrative purposes in detail in
Sec.~\ref{sec:model}.  We note that the choice of potential is not
limited to simple models. In principle any other anisotropic complex
potential can be used, even an anisotropic  protein-protein interaction  derived from all atom MD simulations. 

The particles are propagated  with the first order Brown integrator
explained in Algorithm 1:

\begin{breakablealgorithm}
  \caption{The Brownian dynamics integrator \cite{Davidchack} used in
    our multi-scale scheme. We consider $n$ particles in three
    dimensions 
    with center of mass coordinates
    $\mathbf{r}=(r^{1\top},\dots,  r^{n\top})^\top\in\mathbb{R}^{3n}$,
    $r^j=(r_1^j,r_2^j,r_3^j)^\top \in \mathbb{R}^{3}$, and rotational
    coordinates in the quaternion representation
    $\mathbf{q}=(q^{1\top},\dots,q^{n\top})^\top,
    q^j=(q_0^j,q_1^j,q_2^j,q_3^j)^\top \in \mathbb{S}^3 $, such that
    $|q^j|=1$. 
  Particles are characterized by their
    mass $m$, the mass moment of inertia  $M=\frac{8}{15}m\sigma^2$, and the
  translational and rotational friction coefficients,  $\gamma$ and
    $\Gamma$, respectively.  Note 
    these parameters can differ among species.  Furthermore,    
$\delta t$ is the time-step used in the simulations,
 $\beta=\frac{1}{k_{\rm B}T}$,
$\xi_k$ and
    $\eta_k^{j,l}$ are independent and identically distributed (i.i.d.) Gaussian random variables. $\mathbf{f}$
    is the total force and $\mathbb{F}$ is the total torque, which
    follow from the interaction potential. Finally,  we define  three $4\times 4$ matrices
{$S_1= \left[  \begin{smallmatrix} 0 & -1 & 0 & 0 \\ 1 & 0 & 0 & 0 \\      0 & 0 & 0 & 1\\ 0 & 0 & -1 & 0  \end{smallmatrix}\right],
S_2= \left[   \begin{smallmatrix} 0 & 0 & -1 & 0 \\ 0 & 0 & 0 & -1 \\ 1 & 0
    & 0 & 0 \\ 0 & 1 & 0 & 0  \end{smallmatrix}\right],
S_3= \left[   \begin{smallmatrix} 0 & 0 & 0 & -1 \\ 0 & 0 & 1 & 0 \\ 0 & -1 & 0 & 0 \\ 1 & 0 & 0 & 0  \end{smallmatrix}\right]
$ }.}  
\label{alg:alg1}
 \begin{algorithmic}
 \STATE {{$\mathbf {R_0 = r}$}, {$\mathbf {Q_0 = q}$}, $|q^j|=1$, $j=1,\dots,n$,} \\
 \STATE { {${\bf R}_{k+1} = {\bf R}_{k}+\frac{\delta t}{\gamma m} 
       {\bf f}({\bf R}_k,{\bf Q}_k)+ \sqrt{\delta t}\sqrt{\frac{2}{\gamma\beta m}}\xi_k$,}  }
  \STATE { {$$Y_{k}^j = \frac{\delta t}{\Gamma M} {\bf \mathbb{F}}_j({\bf
        R}_k,{\bf Q}_k)+ \sqrt{\delta t}\sqrt{\frac{2}{\Gamma\beta M}}\sum_{l=1}^{3}\eta_k^{j,l}S_l,$$ }  }\\
  \STATE {$Q_{k+1}^j = \textrm{exp}(Y_{k}^j)Q_k^j$}\\
\end{algorithmic}
\end{breakablealgorithm}

\subsection{Green's functions for rotations}

GFRD handles the free diffusion of single particles.  A freely-moving
particle will undergo rotational as well as translational diffusion.
Although the interactions between particles are anisotropic, we model
the particles as spheres of finite radius for the purpose of modeling
diffusion. This assumption allows the decoupling of the rotational and
translational diffusion of isolated particles, which is possible since
in MD-GFRD the GFRD domains only contain single particles. The Green's
functions for translational diffusion are given by the Green's
functions for single particles inside Single GFRD domains, detailed in
previous work \cite{Vijaykumar2015}. These Green's functions determine
(probabilistically) when the
particles escape from their respective domains, or what their radial
positions inside the domains become when the domains are burst.
Although rotational motion does not influence the center-of-mass
dynamics of a freely diffusing particle, and hence cannot cause escape
from Single Domains, it is nonetheless important to reproduce the
decorrelation of orientations for particles evolving under GFRD. For
example, simply drawing orientations at random when a particle leaves
a GFRD Single Domain will lead to unphysically rapid decorrelation of
orientations when domains are short-lived, and influence properties
such as rebinding probability.

  More
  specifically, on bursting or escape from a Single Domain,  a new orientation $\Omega$ is drawn using
the Green's function $G(\Omega, \Omega_0, t)$, with $\Omega_0$ being
the initial orientation and $t$ the time since domain formation. The
Green's functions, expressed in terms of Euler angles $\alpha, \beta,
\gamma$, can be found in the
literature~\cite{Loman2010,Versmold1970,Favro1960}. For particles with
spherically symmetric diffusion tensors the Green's function is
\begin{widetext}
\begin{equation}
G(\alpha, \beta, \gamma, \alpha_0, \beta_0, \gamma_0, t) = \sum_{L=0}^{\infty} \sum_{K,M=-L}^L \frac{2L+1}{8 \pi^2}  D_{K,M}^{(L)*}( \alpha_0, \beta_0, \gamma_0) D_{K,M}^{(L)}( \alpha, \beta, \gamma) \exp(-D_rL(L+1)t).
\label{eq:rot-GF}
\end{equation}
\end{widetext}
Here, $D_r$ is the threefold degenerate eigenvalue of the diffusion
tensor, given by $D_r= k_{\rm B}T/(8\pi \eta R^3)$ for a particle
of radius $R$ in a fluid of viscosity $\eta$. The quantities $
D_{K,M}^{(L)}( \alpha, \beta, \gamma)$ and its complex conjugate $
D_{K,M}^{(L)*}( \alpha, \beta, \gamma)$ are elements of the Wigner
rotation matrices \cite{Loman2010,Versmold1970,Favro1960}:
\begin{equation}
D_{K,M}^{(L)} ( \alpha, \beta, \gamma) = \exp(-iK\alpha) \,d_{K,M}^{(L)} (\beta)  \exp(-iL\gamma),
\end{equation}
with 
\begin{widetext}
\begin{align}
d_{K,M}^{(L)} (\beta)  = & \left( (L+K)!(L-K)!(L+M)!(L-M)! \right)^{1/2}  \times 
  \\ &  \sum_{S=\max(0,M-K)}^{\min(L+M, L-K)}  
\bigg( \frac{(-1)^{K-M+S}}{(L+M-S)!S!(K-M+S)!(L-K-S)!}  
[\cos(\beta/2)]^{2L+M-K-2S}  [\sin(\beta/2)]^{K-M+2S} \bigg). \nonumber
\label{eq:small_d}
\end{align}
\end{widetext}
For the purposes of clarity, we emphasize that the Euler angles used here should be understood in the following way. If a body frame $B$ has an orientation  $\Omega = (\alpha,\beta, \gamma)$ with respect to some reference frame $F$, then $B$ can be obtained from $F$ by:
\begin{enumerate}
\item Rotating $F$ around $F_z$ by $\gamma$ to give $F^\prime$.
\item Rotating $F^\prime$ around $F_y$ by $\beta$ to give $F^{\prime \prime}$.
\item Rotating $F^{\prime \prime}$ around $F_z$ by $\alpha$ to give $B$.
\end{enumerate}
Moreover, note that the Green's functions are defined without the Jacobian, so that $(\alpha,\beta, \gamma)$ should be drawn from the distribution $\sin (\beta)G(\alpha, \beta, \gamma, \alpha_0, \beta_0, \gamma_0, t)$.

Drawing directly from such a distribution is non-trivial. However, rejection sampling can be used if the maximum of  $\sin (\beta)G(\alpha, \beta, \gamma, \alpha_0, \beta_0, \gamma_0, t)$ is known. Physically, the most likely orientation is always aligned with the initial direction, which suggests a rejection scheme in which a trial orientation $(\alpha, \beta, \gamma)$ is drawn uniformly from $\left([0,2\pi], [0,\pi], [0,2\pi]\right)$, and accepted with a probability
$$\frac{\sin (\beta)G(\alpha, \beta, \gamma, \alpha_0, \beta_0, \gamma_0, t)}{  \sin (\beta_0)G(\alpha_0, \beta_0, \gamma_0, \alpha_0, \beta_0, \gamma_0, t)},$$ with Euler angles defined with respect to the lab frame. Unfortunately, the angular Jacobian implies that $\sin (\beta)G(\alpha, \beta, \gamma, \alpha_0, \beta_0, \gamma_0, t)$ is not in general maximized by $(\alpha = \alpha_0, \beta = \beta_0, \gamma= \gamma_0)$, violating a requirement of rejection sampling. It is true, however, that $\sin (\beta)G(\alpha, \beta, \gamma, \alpha_0, \beta_0, \gamma_0, t)$ is maximized by $(\alpha = \alpha_0, \beta = \beta_0, \gamma= \gamma_0)$ if $\beta_0 = \pi/2, \alpha_0=0, \gamma_0 =0$. We therefore define a new reference frame $F_{\rm temp}$ for each calculation such that the particle initially has orientation $\Omega_0= (0, \pi/2, 0)$ with respect to $F_{\rm temp}$. Using rejection sampling, we can then obtain a new orientation $\Omega = (\alpha, \beta, \gamma)$ with respect to $F_{\rm temp}$. The particle orientation is updated by first rotating the particle by $-\pi/2$ about the $z$-axis of the original particle frame to obtain a particle aligned with $F_{\rm temp}$, and then performing rotations $(\alpha, \beta, \gamma)$ about the axes of $F_{\rm temp}$ as outlined above.

Even with rejection sampling, drawing from the distribution can be computationally challenging due to the costs of evaluating Green's functions. Eq.\,\ref{eq:rot-GF} has an infinite sum that must be truncated; we perform truncation when new contributions are smaller than the current value by a factor of $10^{8}$. To reduce the cost of the summations, we find it helpful to tabulate factorials. We also note that terms in Eq.\,\ref{eq:rot-GF} can be combined in complex conjugate pairs  to eliminate imaginary numbers during the calculation.

Accurate evaluation of the Green's function is most challenging when
$D_rt < 1$, when  $G(\alpha, \beta, \gamma, \alpha_0, \beta_0,
\gamma_0, t)$ is sharply peaked and many terms are needed. For small
$D_rt$, we use early rejection, discarding a large fraction of draws
of $(\alpha, \beta, \gamma)$ if $(\alpha, \beta-\beta_0, \gamma)$ is
large without evaluating $G(\alpha, \beta, \gamma, \alpha_0, \beta_0,
\gamma_0, t)$, and compensating for this bias at the acceptance
stage. Finally, for values of $D_rt < 0.05$, we use the approximate
approach of rotating about a random axis through an angle $\phi =
\sqrt{\phi_x^2+ \phi_y^2 + \phi_z^2}$, where $\phi_i$ are
i.i.d. random variables drawn from a Gaussian of mean 0 and variance
$2D_rt$ \cite{Furry1957}.

\subsection{Handling the dissociation/association reaction in MD-GFRD}
While particles that are sufficiently far away from each other can be
propagated with GFRD, particles that are within a pre-defined cutoff
distance from each other will be propagated with MD, or, as we restrict
ourselves to here,
BD.  As described in more
detail in the next section, this cut-off distance is beyond the range
of the interaction potential, $r_c$. Indeed, the association between
two particles, which is driven by their inter-molecular attraction
forces, is thus simulated explicitly with BD. Also the dissociation
reaction could in principle be simulated with BD: we could explicitly
simulate the bound monomers in the dimer A-B, until they dissociate
again into A
and B. However, the bound state is typically very stable: the time
the particles spent inside the potential well is typically much
longer than the time it takes for the particles to loose their
orientation and thermalise inside the well. Simulating these particles
explicitly means that much CPU time would be wasted on propagating
them while they simply rattle around each other inside the potential
well. In MD-GFRD, we therefore exploit that dissociation is a rare
event: when two BD particles meet a predefined criterion 
signifying that they are deep inside the interaction well, the two
`reactants' A and B are replaced by species C. In turn, the dissociation of C
into A and B is treated as a first-order reaction ${\rm C} \to
{\rm A} + {\rm B}$ with a dissociation rate constant $k_{\rm d}$.

More specifically, when two BD particles come within a distance such
that their interaction energy $E$ drops below some predefined threshold
$E_{\text{{\text{bind}}}}$, then the particles A and B are
replaced by a single particle of species C, with a position that is given by the
center-of-mass of the reactants A and B.  If space permits, the C
particle is directly put into a GFRD domain, which significantly speeds up the
simulation. If there is no space to construct a protective domain, the
C particle is propagated with BD. The C particle then diffuses, either
explicitly with BD or implicitly with GFRD, until it dissociates again
into
the monomers A and B at a later time $\tau_{\rm d}$. Since the
interaction well is deep, $\tau_{\rm d}$  will be
exponentially distributed:
 \begin{equation}
\label{eq:q_d}
 q_{\text{d}}(t) = k_{\text{d}}e^{ -k_{\text{{\text{d}}}}t }.
\end{equation}
Knowing the dissociation rate constant $k_{\text{{\text{d}}}}$, the
time $\tau_{\rm d}$ can thus be sampled
from
\begin{equation}
 \tau_{\rm d} = -k_{\text{{\text{d}}}}\text{ln}(\mathcal{R}_{\text{{\text{d}}}}),
 \label{eq:dissociation time}
\end{equation}
where $\mathcal{R}_{\text{{\text{d}}}} \in {[0,1]}$ is a uniformly distributed random number.

The intrinsic dissociation rate constant $k_{\rm d}$ could in principle be
inferred from experiments. However, a more consistent and rigorous
approach is to obtain $k_{\rm d}$ from a simulation that is performed
prior to the MD-GFRD simulation of interest. This pre-simulation can
then also be used to generate the distributions of the positions and
orientations of A and B at the moment of dissociation. In the MD-GFRD
simulation, the positions and orientations of the
particles at the moment of dissociation can then be sampled from these
distributions, respectively.

In our previous study on isotropic potentials, we determined $k_{\rm
  d}$ by performing a brute force BD simulation of two particles prior
to the MD-GFRD simulations \cite{Vijaykumar2015}. 
However, the particles in our model interact via an anisotropic interaction
potential. This anisotropic interaction is mediated via patches on the
surfaces of the particles, see \fref{fig:patchyParticle}. The range of
the patch-mediated interaction must be short, in order to provide a
strong anisotropy in the interaction. The short range, however, means
that the well of the patch-mediated potential must be deep in order to
induce significant binding: the depth of the well, $\sim25k_{\rm B}T$,
is much larger than that of isotropic particles, $\sim5k_{\rm
  B}T$. The deep well makes it very hard to obtain good statistics in
determining the distribution of dissociation times via brute force
simulations. However, it is possible to efficiently compute the
dissociation rate with rare event techniques such as Transition
Interface Sampling \cite{vanErp2003} or Forward flux sampling
(FFS)\cite{Allen:2005dn}. Here we use the latter technique, which we
describe in section \ref{sec:ffs}.

\subsection{Coupling BD and GFRD}
Now that we have described how MD-GFRD simulates the association and
dissociation of two particles A and B, we will discuss how the
algorithm switches between BD and GFRD when simulating many particles.
At any one point in time, the simulation consists of a set of isolated
particles inside GFRD domains that each have a radius of at least
$d_{\rm min}$, and a set of particles that are propagated with BD and
interact with each other via a pair potential that has an interaction range
 $r_{\rm c}$. There is also a chronologically ordered next-event list that contains the
times at which the GFRD particles escape from their respective domains,
and the times at which the respective particles dissociate, be they in
GFRD or BD mode. 
The particles that are not inside GFRD domains are propagated with
brute-force BD until the first next-event happens. This event can be
an event from the next-event list, but it can also be the formation of
a GFRD domain or the bursting of a GFRD domain when a BD particle
comes too close it. After the event has been executed, BD propagation
is resumed.

Specifically, before each step of BD propagation, the algorithm checks
for the following events, as illustrated in Fig. \ref{fig:coupling}:

\subsubsection{Escape from a GFRD domain}
\label{sec:DomainEsc}
\begin{figure*}[t!]
\includegraphics[scale=1.1]{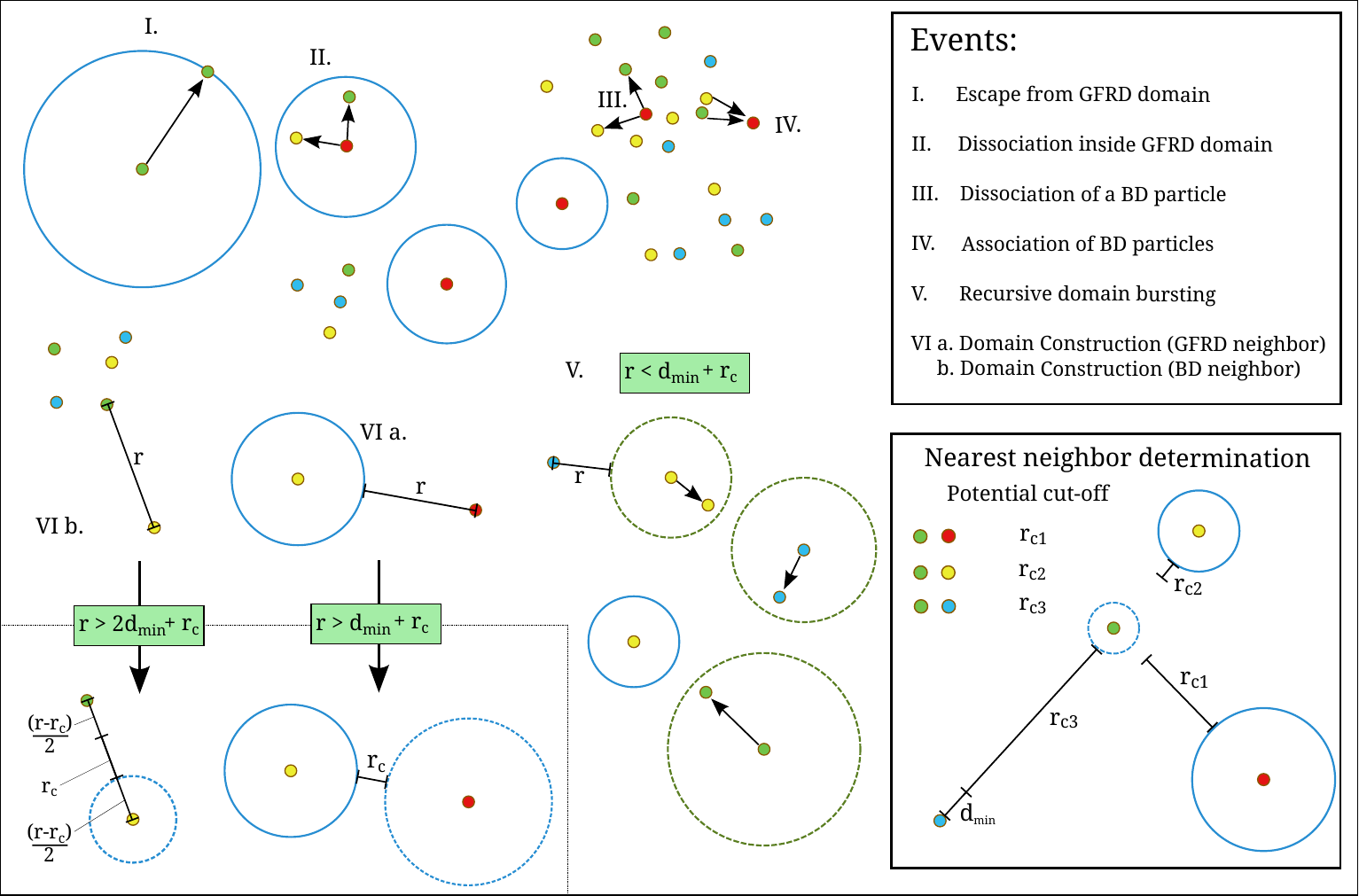}
\caption{At each BD time step, the algorithm  checks whether BD should be
  interrupted. 
The BD propagation is halted when the time
    of the first next event occurs before the global simulation time
    at the end of the time step. 
These next events can be any of the following:
 I. A particle escapes from a
  GFRD domain; the position of the particle is updated to a randomly
  chosen point on the surface of the domain and the domain is
  removed. II. A particle dissociates inside a GFRD domain; the domain
  bursts and the particle is updated to a position and 
  orientation sampled using Green's functions, and is then replaced by
  its product particles. III. A BD particle dissociates; it is
  replaced by its product particles. IV. The binding energy of two
  BD particles is below the binding threshold; the particles enter the bound
  state and are replaced by a single product particle. V. The distance
  from a BD particle to a domain is smaller than $d_{\rm min}+r_{\rm
    c}$; the neighboring domain is burst and the position and
  orientation of the particle in this domain is updated. This particle
  may in turn burst another domain and this happens recursively until
  there is no BD particle within a distance $d_{\rm min}+r_{\rm c}$
  from any other domain. VI.  a. The distance between a BD particle
  and its nearest neighbor is larger than $d_{\rm min}+r_{\rm c}$ in case
  the nearest neighbor is a GFRD domain; a domain of radius $r-r_{\rm
    c}$ is built on the BD particle. b. The distance between a BD
  particle and its nearest neighbor is larger than $2 d_{\rm
    min}+r_{\rm c}$ in case the nearest neighbor is a BD particle; a
  domain of radius $0.5(r-r_{\rm c})$ is built on the BD particle of
  interest.
The inset shows the procedure for
  determining the nearest neighbor, which is the GFRD domain or the BD
  particle with the closest interaction horizon to the (central green) particle of
  interest: for  BD particles the relevant distance is the distance
  minus the sum of the minimum domain radius $d_{\rm min}$ and the
  potential interaction range $r_c$, while for a GFRD domain the
  relevant distance is the distance to the surface of that domain
  minus $r_c$. In the example configuration the blue particle is the
  nearest neighbor. 
}
\label{fig:coupling}
\end{figure*}

When the next event in the list is a particle that escapes from a single domain, that particle is
put at a random center of mass position on the surface of the domain, with an
orientation sampled from Eq. \ref{eq:rot-GF}. The domain is removed
and the particle is put in BD mode. This event is shown in
Fig. \ref{fig:coupling}.I. Note that at the next BD time step, the
algorithm will check whether the particle can be put into a protective
GFRD domain again (see \ref{sec:DomainMake}).

\subsubsection{Dissociation inside a GFRD domain}
\label{sec:GFRDDiss}
When the next event is a particle C inside a GFRD domain decaying into its products  at
time $t$, the domain is burst and a new radial position $r$ for the
reactant is generated according to the normalized translational
Green's function $p(r,t,|r_0,t_0)/S(t-t_0|r_0)$, where $r_0$ is the
original position of the particle, which is the center of the domain
constructed at time $t_0$, and $S(t-t_0|r_0)$ is the survival
probability.  The reactant is replaced by its products, whose
configuration is chosen at random from the ensemble of configurations
recorded at the moment of dissociation, obtained 
 in the FFS pre-simulation. This
event is shown in Fig. \ref{fig:coupling}.II.

\subsubsection{Dissociation of a BD particle}
\label{sec:BDDiss}
When the next event is the dissociation of a BD particle, the particle is replaced by its
products, whose configurations are chosen at random from the ensemble of
configurations recorded at the moment of dissociation in the FFS
pre-simulation. This event is shown in Fig. \ref{fig:coupling}.III.

\subsubsection{Association of BD particles}
\label{sec:BDAss}
When the pair potential energy  between two particles becomes smaller than a
threshold energy, here taken to be $E_{\rm bind} = -10k_{\rm B}T$, the
two particles are defined to be in the bound state.  The two particles
are replaced by a single BD particle at their center of mass. This event is shown in
Fig. \ref{fig:coupling}.IV. Note that at the next BD step, the
algorithm will check whether the particle can be put into a GFRD
domain, as described under \ref{sec:DomainMake}.

\subsubsection{Recursive domain bursting}
\label{sec:DomainBurst}
When a BD particle comes at time $t$ within a distance of $d_{\rm
  min}+r_{\rm c}$ from the surface of a GFRD domain, the domain is
burst and a radial position $r$ of the particle inside that domain is
drawn from the normalized translational Green's function
$p(r,t|r_0,t_0)/S(t-t_0|r_0)$, where $t_0$ is the time and $r_0$ the
position of the center of the domain when it was constructed, and
$S(t-t_0|r_0)$ the survival probability. A new orientation of the
particles is sampled from Eq. \ref{eq:rot-GF}.  If this particle,
after updating its position, comes within a distance of $d_{\rm
  min}+r_{\rm c}$ from another domain, that domain is also burst. This
may lead to a cascade of domain bursting, which ceases when no BD
particle is within a distance of $d_{\rm min}+r_{\rm c}$ from any GFRD
domain. This event is shown in Fig. \ref{fig:coupling}V. Note that
domains are always at least $r_c$  apart from each other.

\subsubsection{Domain Construction}
\label{sec:DomainMake}
For each BD particle, the algorithm determines the nearest neighbor,
which is either another BD particle or a GFRD domain. The procedure to
determine the nearest-neighbor distance depends on whether the
neighbor is a BD particle or a GFRD domain, as shown in the
inset of Fig. \ref{fig:coupling}.  A BD particle is put into a GFRD
domain when the distance $r$  between the particle and its nearest
neighbor:
\begin{enumerate}[label=(\alph*)]
\item is larger than $d_{\rm min}+r_{\rm c}$ in case the nearest
  neighbor is a GFRD domain. A domain of radius $(r-r_{\rm c})$ is
  built around the particle of interest. This event is shown in
  Fig. \ref{fig:coupling}.VI a.
\item is larger than 2$d_{\rm min}+r_{\rm c}$ in case the nearest
  neighbor is a BD particle. A domain of radius $0.5(r-r_{\rm c})$ is
  built around the particle of interest. This allows enough space to
  build a domain with a radius of at least $d_{\rm min}$ around the
  neighbor, thus
  preventing the neighbor from prematurely bursting the newly built
  domain. This event is shown in Fig. \ref{fig:coupling}.VI b.
\end{enumerate}
For the newly constructed domain the tentative next-event times for
the respective tentative event types (e.g. dissociation and escape)
are determined, and the event type with the smallest tentative
next-event time is added to the event list.  To achieve maximum
efficiency, the minimal domain size
$d_{\rm min}$ should be as small as is practical. 


\subsection{Computing the dissociation rate with Forward Flux Sampling\label{sec:ffs}}

\begin{figure}[t]
\includegraphics[width=8cm]{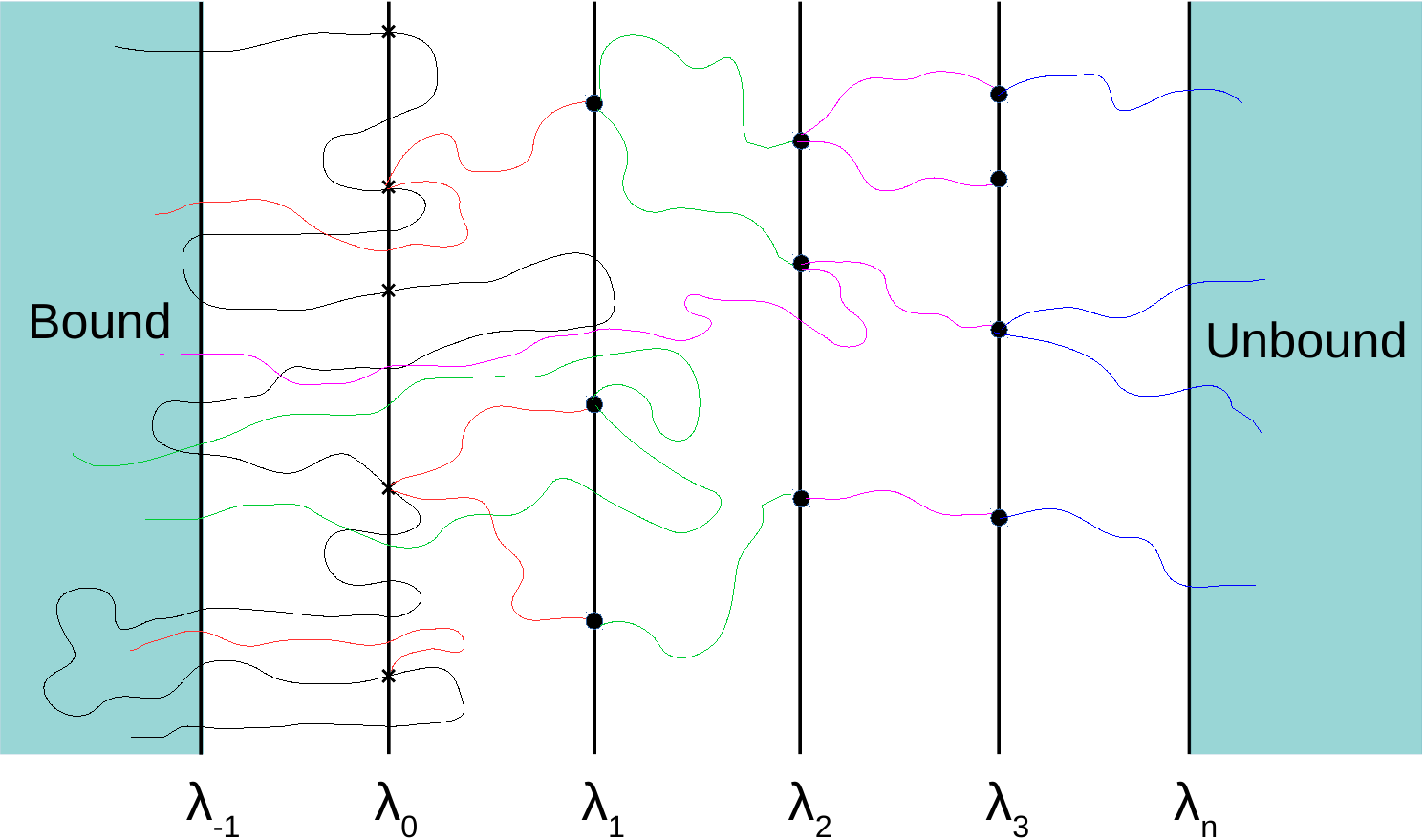}
\caption{An illustration of the FFS method. An ensemble of transition
  paths is generated by starting trial runs from randomly picked
  configurations on interfaces, which are the end points of previous
  successful trial runs.
}
\label{fig:ffs}
\end{figure}

The Forward Flux Sampling (FFS) algorithm enables efficient evaluation of rare event
kinetics. 
FFS uses a series of interfaces between the reactant  and the product
states to construct the transition path ensemble and calculate the
corresponding transition rate. Each interface is defined by an
order parameter $\lambda$: the reactant state 
is defined by
$\lambda < \lambda_{\rm -1}$ and the product state 
by $\lambda > \lambda_n$. The remaining interfaces are
defined by intermediate values of $\lambda$: $(\lambda_0 \dots
\lambda_{{n-1}}$). The FFS technique requires that
$\lambda_{{i+1}} > \lambda_{i}$ for all $i$, and all the
trajectories from reactant to product state
pass through each
interface in succession as shown in Fig. \ref{fig:ffs}.
Trajectories starting in the reactant state and reaching product state
are rare, but trajectories starting at an interface and
crossing the next interface are more common. This is the central idea
used in FFS~\cite{Allen:2005dn}. 

Here we use the `direct' FFS variant, DFFS, to compute the
dissociation rate \cite{Allen:2009kb}. In this process the reactant
state is the bound A,B dimer, and the product state corresponds to
the dissociated dimer.
For the purpose of simulating dissociation, we take an order parameter
to determine the interfaces based on a
combination of the energy of interaction and the inter-particle
distance. 
The reactant bound state interface $\lambda_{\rm -1}$ is defined by a
potential energy $E_{bind}$, while the product state is defined
by zero potential energy in addition to an inter-particle distance larger than the cut-off $r_c$ .

In the first step of FFS, a
  brute-force BD simulation is performed to compute the
  flux $\phi$ of crossing the interface $\lambda_{0}$ while
  coming from the bound state. This brute-force
  simulation generates an ensemble of configurations at
  $\lambda_0$. In the next step, a trajectory is fired from a randomly
  chosen configuration from this ensemble; this trajectory is then
  propagated until it either hits the next interface
  $\lambda_{1}$ or returns to  the reactant state (i.e., recrosses $\lambda_{-1}$).  This procedure is repeated
  until a sufficiently large number of configurations at the next interface
  $\lambda_1$ is generated. The fraction of trajectories that makes it
  from $\lambda_0$ to $\lambda_1$ yields the conditional probability
  $P(\lambda_1|\lambda_0)$ that a trajectory that comes from 
the bound state and crosses $\lambda_0$ for the first time will subsequently
  reach $\lambda_1$ instead of returning to the bound state. This whole
  procedure is then repeated for all subsequent interfaces until the final
  interface $\lambda_n$ is reached, signifying the fully dissociated pair. 
 Under the assumption of rare event kinetics, the intrinsic dissociation rate
$k_{\rm{d}}$ is then given by \cite{vanErp2003,Allen:2005dn}
\begin{equation}
k_{\rm{d}} = \phi \prod_{i=0}^{n-1}P(\lambda_{{i+1}}|\lambda_{i}).
\label{eq:ffskd}
\end{equation}

\subsection{Illustrative anisotropic inter-particle potential}
\label{sec:model}
In this section we describe the interaction potential for the specific
 patchy-particle system. We reiterate
that our multi-scale scheme is independent of the choice of potential,
and can in principle be applied with arbitrarily complex potentials.

For convenience, we split our inter-particle potentials into three
parts. Every pair of particles experiences a repulsive potential
$U_{\rm rep}(R)$ and an isotropic attractive potential $U_{\rm
  isoAtt}(R)$ based on the distance $R$ between the centers of
mass. Additionally, each pair of complementary patches interacts
through an attractive potential $U_{\rm att}(r)$ based on the distance
$r$ between complementary patches (see \fref{fig:patchyParticle}). For
a pair of particles with a single pair of complementary patches,
\begin{equation}
\elabel{eq:Vpot}
U(R,r) = U_{\rm rep}(R) +  U_{\rm isoAtt}(R) + U_{\rm att}(r).
\end{equation}
Mediating the attractive interactions through surface-based patches
naturally captures short-range contact interactions.

It is common to use 12-6 Lennard-Jones or related potentials in
biomolecular modeling.  Although the $r^{-6}$ dependence is required
for van der Waals interactions between atoms and even between larger
entities, in general there is no fundamental reason to choose this
functional form in case of complex effective interactions between
biomolecules, e.g. hydrophobic interactions.  In preliminary
simulations, we observed that using Lennard-Jones potentials leads to
numerical difficulties, forcing the use of extremely small time
steps. The underlying reason is that Lennard-Jones potentials have a
large curvature close to the minimum of the bound state, a situation
for which the Brownian integrator is poorly suited. This effect is
exacerbated by the use of short-ranged anisotropic attractions
between particles, which reduces the entropy of the bound state and
must be compensated for by stronger attractive potentials, in order to
model realistic equilibrium binding constants.  Stronger attractive
potentials lead to larger second derivatives of the
potential. Moreover, requiring potentials to be short-ranged and
orientation-specific implies variation over short length and angular
scales, again increasing the second derivatives of the potential.

\begin{figure}[b]
\includegraphics[width=8.3cm]{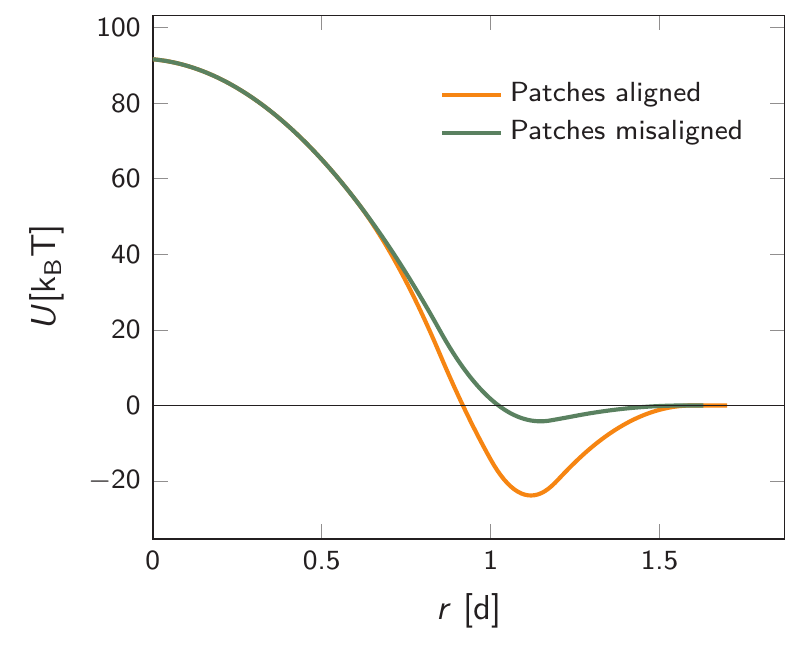}
\caption{Inter-particle interactions. Total interaction potential $U_{\rm rep}(R)+U_{\rm isoAtt}(R)+U_{\rm att}(R-2d_{\rm patch})$ for two particles with perfectly aligned complementary patches, and the total interaction potential $U_{\rm rep}(R)+U_{\rm isoAtt}(R)+U_{\rm att}(R+2d_{\rm patch})$ when the complementary patches are completely misaligned. The existence of patches introduces an attractive bound state with the particles in close contact.}
\label{fig:Pot-1d}
\end{figure}

\begin{figure*}[t]
\includegraphics[width=16.6cm]{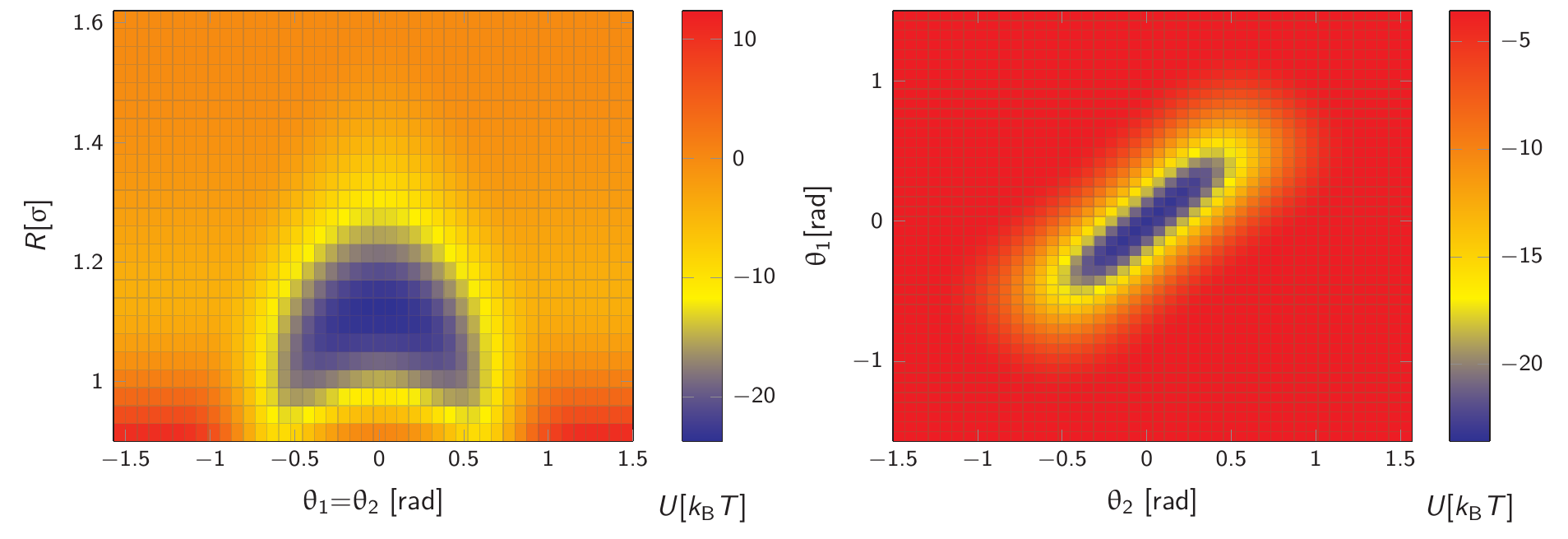}
\caption{Total potential $U_{\rm rep}(R)+U_{\rm isoAtt}(R)+U_{\rm
    att}(r)$ for two particles with complementary patches.  (a) Potential as a
  function of the distance between centres of mass $R$ and alignment
  of patches with inter-particle vector, $\theta_1$ and $\theta_2$
  (see definition in \fref{fig:patchyParticle}), given
  $\theta_1=\theta_2$. (b) Potential as a function of
  $\theta_1,\theta_2$ given $R=1.1\sigma$. Note the relatively narrow
  range of orientations over which strong bonding occurs. }
\label{fig:Pot-2d}
\end{figure*}

Instead of using a Lennard-Jones type potential we therefore illustrate our method using piece-wise quadratic potentials  similar to those employed elsewhere \cite{Ouldridge2011}. These potentials give us more control over the shape, and allow for easier integration with potentials that are short-ranged and highly orientation-specific. We stress that using an alternative potential that is more challenging for the integrator would not remove the advantages of the multi-scale scheme.

  $U_{\rm rep}(R)$, $U_{\rm isoAtt}(R)$ and $U_{\rm att}(r)$ have the form
\begin{equation}
\elabel{pot}
U_i(x) = \begin{cases}
	 \epsilon_i (1 - a_i \left(\frac{x}{\sigma}\right)^2)  & \text{if $ x <  x^{\star}_i $},\\
	 \epsilon_i b_i(\frac{x^c_i}{\sigma} - \frac{x}{\sigma})^2 & \text{if $x^{\star}_i < x <x^c_i$},\\
	0 & \text{otherwise},
	\end{cases} 
\end{equation}
with $i={\rm rep}, {\rm isoAtt}, {\rm att}$, respectively.  The
overall strength $\epsilon_i$, the length scale $\sigma$ (i.e.~the particle
diameter), the
stiffness $a_i$ and the parameter $x^{\star}_i $, which combined with
$a_i$ determines the range of the potential, are free
parameters. Cut-offs $x^c_i$ and smoothing parameters $b_i$ are fixed
by requiring continuity and differentiability at $x^\star_i$. For our
illustrative purposes, we take the following parameters:
$\epsilon_{\rm rep}=100k_{\rm B T}$, $a_{\rm rep}=1$ and $R^*_{\rm
  rep}=0.85\sigma$, implying $b_{\rm rep}=2.6036$ and $R^c_{\rm
  rep}=1.1764\sigma$; $\epsilon_{\rm att}=20k_{\rm B}T$, $a_{\rm att}=20$
and $r^*_{\rm att}=0.1\sigma$, implying $b_{\rm att}=5$ and $r^c_{\rm
  att}=0.5\sigma$; and $\epsilon_{\rm isoAtt}=10k_{\rm B T}$, $a_{\rm
  isoAtt}=1$ and $R^*_{\rm isoAtt}=0.85\sigma$, implying $b_{\rm
  rep}=2.6036$ and $R^c_{\rm rep}=1.1764\sigma$. 

In Fig. \ref{fig:Pot-1d}, we plot the resulting total inter-particle
potential as a function of distance $R$ when the two complementary patches are
aligned to face each other, so that $r= R-\sigma$. A narrow attractive well corresponding to the
  two particles being in close contact is evident. For comparison, we also show the total inter-particle potential
  as a function of $R$ when the two complementary patches are
  misaligned to face opposite each other, so that $r= R+\sigma$. In this case, the patches do not contribute to the
    interaction; the non-specific, isotropic part of the potential,
    however, still gives rise to a weak attraction.  In Fig.\,\ref{fig:Pot-2d}, we demonstrate the
  orientational dependence of the attractive potential, showing that
  the attractive interaction is highly sensitive to misalignment.  We
  note that our choice of potential makes truncation at short
  distances relatively trivial. This is helpful in allowing rapid
  switching to GFRD domains once the particles are separated.
  
For our model potential the
interaction range  is set $r_{\rm c}=1.6\sigma$, where the pair
potential in \eref{eq:Vpot}
has vanished. Moreover, in the
MD-GFRD simulations we set the minimum domain size
$d_{\rm min}=0.5\sigma$ and the  particle diameter to  $\sigma=5$nm.


\section{Results and discussion}	
We test the MD-GFRD simulation using the patchy-particle model
described in section \ref{sec:model}.  In the simulations there
are three species of particles, A, B and C, which react according to
\begin{equation}
A+B \leftrightarrows C.
\label{eq:CR}
\end{equation}
The system specific parameters of the simulation are as follows:
The particle diameter is $\sigma=5$nm, the time step $\delta t = 0.1 $ns,
the mass of the particle is $m=50$kDa, the mass moment of inertia $M=\frac{8}{15}m\sigma^2$ the translational and rotational diffusion
constants, are $D_t = 1\rm{\mu m^2/s}$ and $D_r =
1.6\times10^{7}\rm{rad^2/s}$ for all particles, the translational and rotational friction coefficients are $\gamma=\frac{k_{\rm B}T}{D_tm}$
the $\Gamma=\frac{k_{\rm B}T}{D_rM}$ respectively, where $k_{\rm B} = 1.38\times10^{-23}\rm{JK^{-1}}$ is the Boltzmann constant and $T=300$K is the temperature of the system.
In the following subsections we first
present the results of the FFS-BD pre-simulation used to determine the
value of the intrinsic dissociation rate $k_{\rm d}$. Next, using the value of $k_{\rm d}$, we perform
MD-GFRD simulations in which we compute the probability that A and B
are bound, as a function of system size. We compare the results
against Monte Carlo simulations and analytical
expressions. We compute the power spectra for the binding
process. Finally, we discuss  the performance of the algorithm.

\subsection{Rate constant determination using FFS-BD pre-simulation}

\begin{figure}[b]
\includegraphics[width=8.3cm]{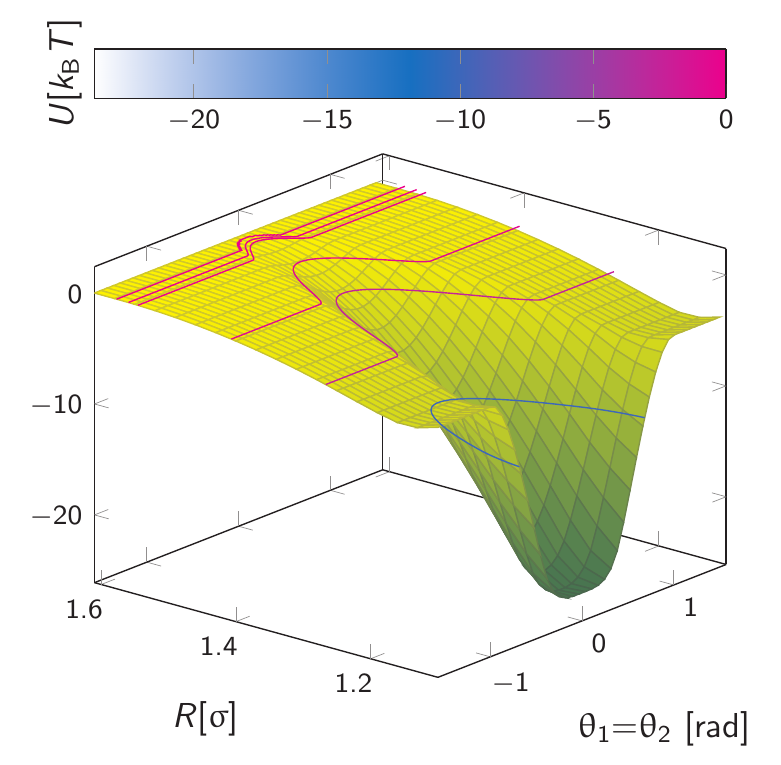}
\caption{FFS interfaces were defined by the  potential
  energy:  $\lambda_0 =
-10 k_{\rm B}T$,  $\lambda_1 =-2.5 k_{\rm B}T$,  $\lambda_2 =-0.75
k_{\rm B}T$,  $\lambda_3 =-0.025 k_{\rm B}T$,  $\lambda_4 =-0.0075
k_{\rm B}T$. The final interface $\lambda_5$ was defined by zero
energy and a distance $R>1.6 \sigma$.
Using these interfaces as starting points for successive trial runs, the particles are driven from the bound to the unbound state.}
\label{fig:ffs2}
\end{figure}

As explained in Sec. \ref{sec:ffs}, it is advantageous to treat the
dimer A-B  as a single particle C, which then can dissociate again into A and B
with an intrinsic rate $k_{\rm d}$. We used direct FFS to precompute the intrinsic  rate
constant $k_d$. The interfaces $\lambda_i$ are defined in terms of the
interaction energy, as shown in Fig. \ref{fig:ffs2}.
The bound state interface  $\lambda_{-1}$ was  defined by $U(R,r) < -10 k_{\rm B}T$,
the dissociated  state final interface $\lambda_5$ was set at a distance $R= 1.6
\sigma$. Five intermediate interface were set at respectively  
 $\lambda_0 =
-10 k_{\rm B}T$,  $\lambda_1 =-2.5 k_{\rm B}T$,  $\lambda_2 =-0.75
k_{\rm B}T$,  $\lambda_3 =-0.025 k_{\rm B}T$,  $\lambda_4 =-0.0075
k_{\rm B}T$.
A straightforward BD trajectory created 100,000 configurations at the first
interface. Subsequently, performing direct FFS yielded
20,000 configurations for each successive interface. Using
Eq. \ref{eq:ffskd}, we find for the intrinsic dissociation rate
constant $k_{\rm d} = 4.66 \rm s^{-1}$. The configurations at the final interface can be used to draw from when
performing the dissociation step in the MD-GFRD, see Sec.~\ref{sec:GFRDDiss}.

\subsection{Bimolecular reactions}

To test the multi-scale scheme, we simulate the bi-molecular reaction
shown in Eq. \ref{eq:CR}. In these simulations we start off with two
species of particles A and B, each having one patch on its surface. An
A particle can react with a B particle to form a dimer. Also, a C
particle can dissociate to form one A and one B, with an intrinsic rate $k_{\rm
  d}$ that has been pre-computed using FFS (see previous section).  We
assume that the mixture is ideal: only species A and B have an
attractive interaction $U(R,r)$. All other interaction potentials
between pairs A-A, B-B, C-C, C-A and C-B are repulsive only.  We test the scheme for two different scenarios, one
starting with a single A and a single B particle and the second
starting with two A and two B particles. The simulation results are
compared with Monte Carlo simulations of the
same model and with analytical expressions.

In the first case, one particle of species A and one particle of
species B, each having one patch, are put in a cubic  box of volume $V$, with
periodic boundary conditions.  This means that the number of C
particles, $N_C$, is either zero or one. From the computed time
average of $N_C$, we calculate the probability $\mathcal{P}_b$ that
the A particle is bound to B. We repeat this procedure for different
box sizes.  In
\fref{fig:pbOneAOneB} we compare the value of $\mathcal{P}_b$ obtained
using the new MD-GFRD algorithm to the results obtained from Monte
Carlo simulations of the same system. The figure also shows the
analytical result
\begin{equation}
 \mathcal{P}_b = \cfrac{\langle N_C\rangle}{N_A} = \frac{k_{\rm on}}{k_{\rm on}+Vk_{\rm off}} = \frac{\phi(V)}{\phi(V)+1},
 \label{eq:1A1B}
\end{equation}
where $\langle N_C\rangle$ is the average of $N_C$
 and
$\phi(V)$ is the ratio of the probability that an A particle is
bound versus unbound
\begin{equation}
\phi(V)= \frac{k_{\rm on}}{Vk_{\rm off}} = \frac{K_{\rm eq}}{V}.
\label{eq:phi_V}
\end{equation}
Here, $k_{\rm on}$ and $k_{\rm off}$ are the effective association and
dissociation rates, respectively, and $K_{\rm eq}$ is the equilibrium
constant 
\begin{equation}
 K_{\rm eq} = \int d{\bf R} \int d \hat{\bf u}_1 \int d\hat{\bf u}_2
 e^{-\beta V(R, r({\bf R},\hat{\bf u}_1, \hat{\bf u}_2))},
 \label{eq:Keq}
\end{equation}
where $U(R, r({\bf R},\hat{\bf u}_1 \hat{\bf u}_2))$ is
the interaction potential given by \eref{eq:Vpot}, with 
${\bf R}$ the inter-particle vector, $R$ the magnitude of ${\bf R}$, $r$ the distance between the patches of
the particles, which depends on ${\bf R}$ and the orientation of the
two particles denoted by the patch vectors in the stationary lab frame, $\hat{\bf u}_1$ and $\hat{\bf u}_2$,
respectively given by Eq. \ref{eq:spacePatch}. Solving Eq. \ref{eq:Keq} analytically is not possible
for the complex anisotropic potential used here. However, recently we
have shown how in one TIS/FFS simulation both the association rate
$k_{\rm on}$ and the
dissociation rate $k_{\rm off}$ can be computed \cite{vijaykumarFD}, which then allows us
to obtain $K_{\rm eq}$ from Eq. \ref{eq:phi_V}. Applying this
technique to this potential revealed that $k_{\rm on}=0.135\mu {\rm
  m^3}{\rm s}^{-1}$ and $k_{\rm off}=1.384{\rm
  s}^{-1}$. \fref{fig:pbOneAOneB} shows that the results of the
MD-GFRD simulations agree very well with both the results of the Monte
Carlo simulations and with the analytical predictions.

\begin{figure}
\includegraphics[width=8.3cm]{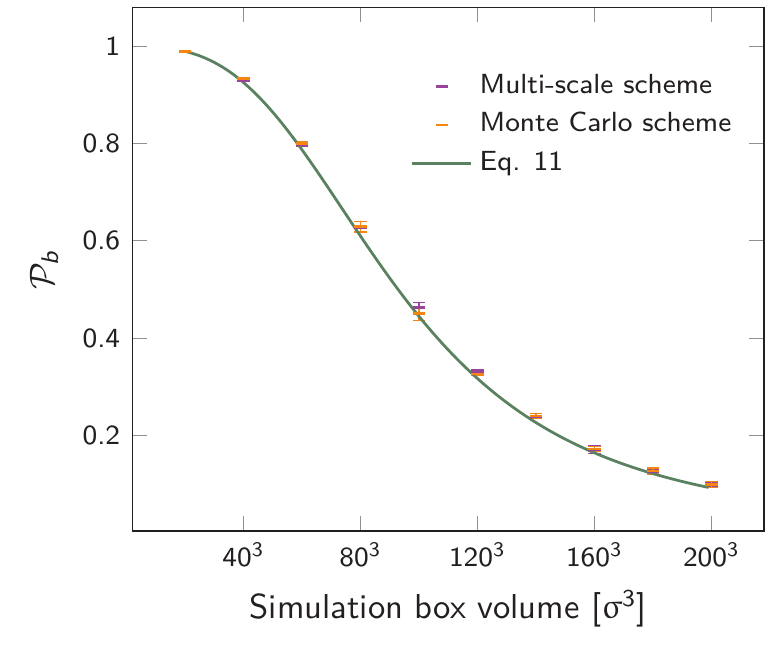}
\caption{The probability $\mathcal{P}_b$ that a particle A is bound to
  a particle B, as a function of the volume of the box.
  Simulations are performed with one A particle and one B particle in
  the box. The points with the error bars are the results of the
  MD-GFRD simulations and the Monte Carlo simulation. These
  results are validated with the analytical prediction of
  Eq. \ref{eq:1A1B}. It is seen that the agreement is very good. The
  translational and rotational diffusion constants, which are not
  important for the value of $\mathcal{P}_b$, are $1\rm{\mu m^2/s}$
  and $1.6\times10^{7}\rm{rad^2/s}$ .}
\flabel{fig:pbOneAOneB}
\end{figure}

\begin{figure}
\includegraphics[width=8.3cm]{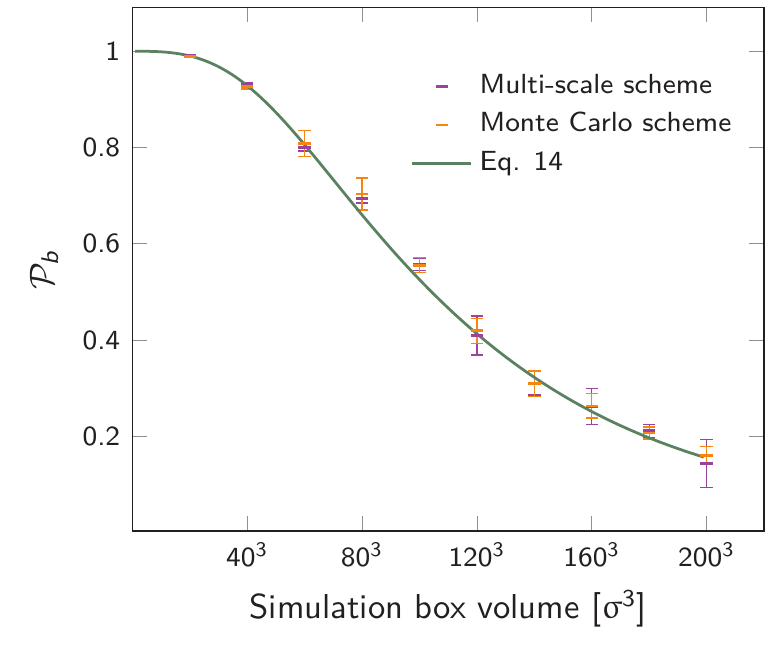}
\caption{The probability $\mathcal{P}_b$ that a particle A is bound to
  a particle B, as a function of the volume of the box.
  Simulations are performed starting with two A particles and two B particles in the
  box. The points with the error bars show the results of the new
  MD-GFRD scheme and the Monte Carlo simulations. These results are validated with the analytical prediction of Eq. \ref{eq:2A2B}. It is seen that the agreement is very
  good. The translational and rotational diffusion constants, which are not
  important for the value of $\mathcal{P}_b$, are $1\rm{\mu m^2/s}$ and $1.6\times10^{7}\rm{rad^2/s}$ .}
\flabel{fig:pb2A2B}
\end{figure}

In the second test, we start with 2 A particles and 2 B particles,
which can again interact via the same interaction potential to form
species C. We can analytically compute the probability that an A
particle is bound to a B particle, by carefully summing over all
possible configurations \cite{Tom2010}:
\begin{equation}
 \mathcal{P}_b = \frac{\phi(V)+\phi(V)^2}{2(0.25+\phi(V)+\frac{\phi(V)^2}{2})},
 \label{eq:2A2B}
\end{equation}
where $\phi(V)$ is given by Eq. \ref{eq:phi_V}. The results of the MD-GFRD
simulations, the Monte Carlo simulations, and the analytical
prediction are shown in \fref{fig:pb2A2B}. It is seen that the agreement
is very good.

\subsection{Power Spectrum}
\begin{figure}[t!]
\includegraphics[width=8.3cm]{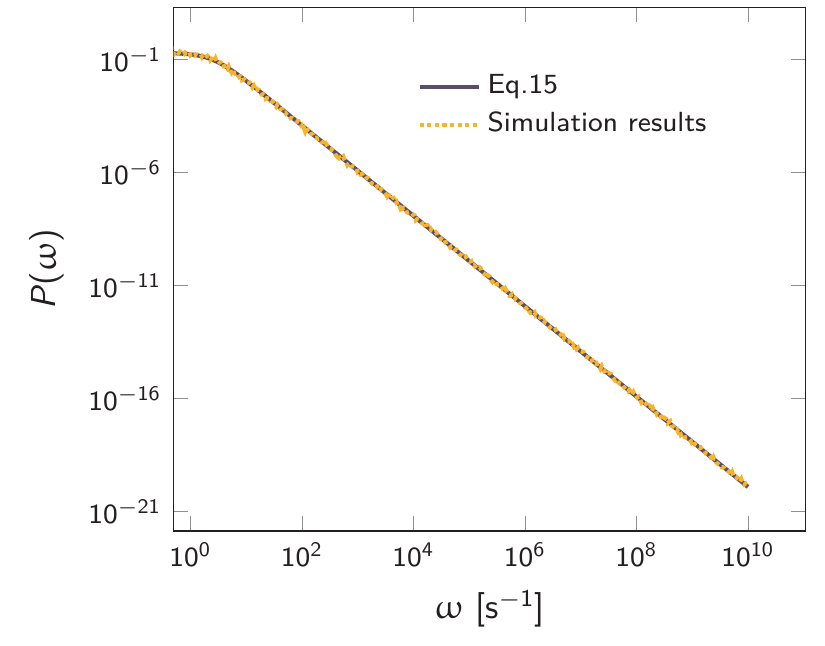}
\caption{MD-GFRD successfully predicts the power spectrum
  $P_n(\omega)$ of the binding state $n(t)$ of two particles
  switching between the bound state $n(t)=1$ and the unbound state
  $n(t)=0$. The dotted line shows the results of the MD-GFRD
  simulations, while the solid line shows the analytical prediction of
  Eq. \ref{eq:ps}, where the association rate $k_{\rm on}$ and
  dissociation rate $k_{\rm off}$ have been computed from a single FFS
  simulation as described in Ref. \cite{vijaykumarFD}. Two particles, one of each
  species A and B were simulated in a box of side length 100$\sigma$.}
\label{fig:ps}
\end{figure}

We can use MD-GFRD to compute the power spectrum $P_n(\omega)$ of the
time trace of the binding state $n(t)$ of two particles, switching
between the bound state with $n(t)=1$ and the unbound state with
$n(t)=0$. The dotted line in Fig. \ref{fig:ps} shows the result. We
expect that this power spectrum is given by that of a random telegraph
process \cite{Kaizu:2014eb}:
\begin{equation}
P(\omega) = \frac{2\mu \mathcal{P_{\rm b}}(1-\mathcal{P_{\rm b}})}{\mu^2+\omega^2},
\label{eq:ps}
\end{equation}
where $\omega$ is the frequency, $\mu=k_{\rm on}/V+k_{\rm off}$ is the renormalized/effective
decay rate, and $\mathcal{P_{\rm b}}=k_{\rm on}/(k_{\rm on}+Vk_{\rm
  off})$ is the binding probability.  To predict the power spectrum,
we thus need the effective association rate $k_{\rm on}$ and the
effective dissociation rate $k_{\rm off}$. As described in the
previous section, these rates can be computed in a single TIS/FFS
simulation \cite{vijaykumarFD}. Using the computed values of the rate
constants in combination with Eq. \ref{eq:ps}, we arrive at the
analytical prediction of the solid line in Fig. \ref{fig:ps}. It is
seen that the agreement with the MD-GFRD simulation results is
excellent.  MD-GFRD thus not only reproduces mean quantities but
also successfully predicts dynamic quantities.

\subsection{Performance}
\begin{figure}[t!]
\includegraphics[width=8.3cm]{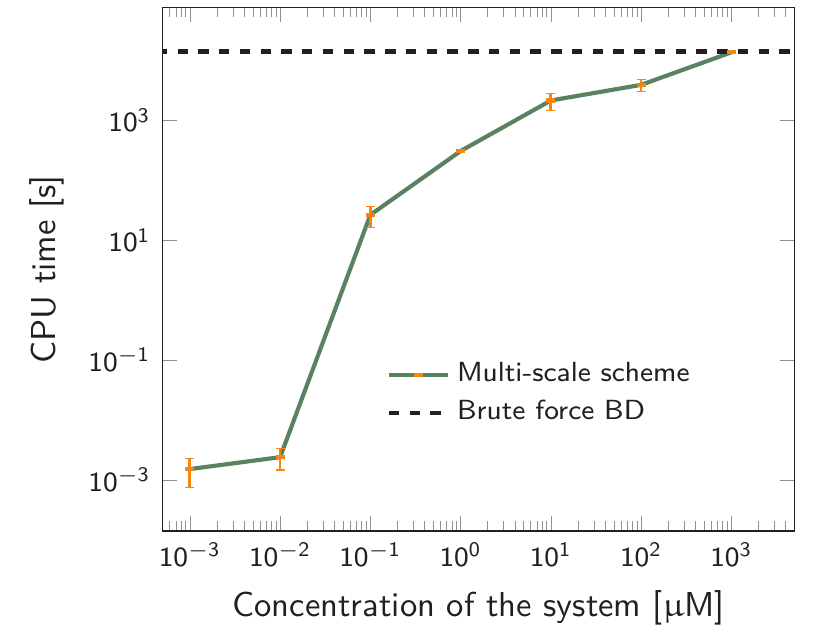}
\caption{The CPU time to simulate 1ms real time as a function of the
  concentration of A and B, for MD-GFRD (solid line) and BD (dashed line).  The concentration is
  varied by changing the volume of the simulation box, while the
  number of particles is kept constant at $N_A=N_B=5$. 
  It is seen that in the  biologically relevant concentration range of 
  nanomolar to micromolar
  the performance of MD-GFRD is much better than
  that of brute-force BD, but at higher concentrations the
  relative performance of MD-GFRD goes down. This is because at higher
  concentrations, the particles will be close to each other, and the
  system cannot capitalize on the potential of MD-GFRD to make large
  jumps in time and space.}
\label{fig:perf}
\end{figure}

The motivation to combine GFRD and MD into a multi-scale scheme is
  the computational speed up it can provide. Unlike brute force
  Brownian dynamics which spends a lot of CPU time in propagating the
  particles toward each other, GFRD makes large jumps in space and
  time when the particles are far apart from each other and the GFRD
  domains are large.  The
  computational power of GFRD can thus especially be reaped when the
  particles are often far apart, which is the case when the
  concentrations are low. This can be seen in Fig. \ref{fig:perf}, which shows a
  comparison of MD-GFRD against brute force BD as a function of
  concentration. It is seen that MD-GFRD is much more efficient than
  brute force BD, especially when the concentrations are below a
  $\mu{\rm M}$. However, for high concentrations, the performance of
  MD-GFRD becomes comparable to that of BD. In this regime, the
  particles are often so close together that no big jumps in time
  and space can be made. Interestingly, however, the crossover happens
  only at a mM concentration, which means that for most biologically
  relevant concentrations MD-GFRD is much faster than brute-force BD.

\section{Conclusion}

In this work we extended the MD-GFRD scheme
\cite{Vijaykumar2015} 
to include the orientational
  dynamics of the particles, enabling the 
simulation  of 
 reaction and diffusion of particles that interact
  via anisotropic interaction potentials. This opens up the possibility
  to treat a whole class of
  interesting problems. Biomolecules such as proteins and DNA
  typically interact with each other via anisotropic potentials. In
  some cases of biological interest the dynamics at short length and
  time scales can be integrated out
\cite{VanZon:2006bn,Morelli:2011ie,Mugler:2013wx,
  tenWolde:2016ih}. For example, a gene regulatory protein that
  has just dissociated from its promoter on the DNA either rapidly
  rebinds the DNA or rapidly escapes into the bulk, where it will
  loose its orientation; conversely, a new protein tends to arrive at
  the promoter from the bulk in a random orientation. In these cases,
  we expect that the regulatory proteins can be modeled as isotropic
  particles that interact with the DNA via effective rate constants,
  which take into account the anisotropy of the interaction. However,
  it is now well established that in many systems the dynamics at
  molecular length and timescales, arising from e.g. enzyme-substrate
  rebindings, can qualitatively change the macroscopic behavior of the
  system at cellular length scales
\cite{Takahashi2010,Aokia:2011ji}. This phenomenon can occur in
  biochemical networks with multi-site protein modification, which are
  omnipresent in cellular biology \cite{Takahashi2010}. In such
  systems, the orientational dynamics cannot be integrated out: the
  probability that an enzyme which has dissociated from its substrate
  molecule rebinds to another site on the same substrate molecule to
  chemically modify it, will depend in a non-trivial manner on the
  translational and orientational diffusion constants of the
  particles, their size, and the distance between the patches on the
  substrate. The MD-GFRD scheme presented here now makes it possible
  to study the interplay between the microscopic dynamics at the
  molecular scale and the network dynamics at the cellular scale in
  this large class of systems.

In addition, the MD-GFRD scheme could more generally be used for
soft matter self-assembly
where building blocks that are diffusing in the dilute solution come
together and bind occasionally to form large complexes and structures\cite{Glotzer:2007}.

\section{Acknowledgements}
 This work is part of the Industrial Partnership Programme (IPP)
    `Computational sciences for energy research' of the Foundation for
    Fundamental Research on Matter (FOM), which is financially
    supported by the Netherlands Organization for Scientific Research
    (NWO). This research programme is co-financed by Shell Global
    Solutions International B.V.


\bibliography{library.bib}

\end{document}